\documentstyle[aps,prb,preprint]{revtex}

\begin{document}
\author{O.V. Danylenko$^{1}$ and O.V. Dolgov$^{2}$}
\address{$^1$ Institut f\"ur Theoretische Physik, Universit\"at T\"ubingen, \\
Auf der Morgenstelle 14, D-72076 T\"ubingen, Germany \\
$^2$ Max-Planck-Institut f\"ur Festk\"orperforschung,\\
Heisenbergstr. 1, D-70506 Stuttgart, Germany}
\title{Nonadiabatic Contribution to the Quasiparticle Self-Energy in Systems with
Strong Electron-Phonon Interaction}
\date{}
\maketitle

\begin{abstract}
We investigate effects of a nonadiabatic electron-phonon(boson) interaction
on the quasiparticle self-energy in the lowest order in the coupling
constant. Existing approaches either overestimate, or underestimate these
effects because of different approximations for momentum and frequency
dependences of the vertex corrections. The connection between the
nonadiabaticity and a possible instability of the interacting Fermi system
is discussed as well.
\end{abstract}

\section{Introduction}

\label{SEintroductionpage} The standard description of electron-phonon
interaction (EPI) in metals is based on the{\sc \ }so-called
Born-Oppenheimer \cite{BO}( or adiabatic) theorem. Its main statement says
that electrons are not sensitive to the{\sc \ }motion of ions and are
influenced only by their static electric field. One can to say that ''fast''
electrons, due to electroneutrality, follow ''slow'' ions. Narrow bands and a%
{\sc \ }strong electron-phonon interaction make their velocities comparable
in order of magnitude. An appropriate mathematical description is the{\sc \ }%
standard Feynman-Dyson perturbation theory, where phonons are considered to
be uneffected by the{\sc \ }EPI. However, there are some processes which in
higher orders violate this approximation. Mathematically this can be taken
into account by virtue of the so-called ''vertex corrections''. These
effects were considered by Migdal \cite{M} who showed that their
contributions are small,{\sc \ }of the order of a small parameter $\Omega
_{ph}/\tilde{W}$, where $\tilde{W}$ is a characteric energy, defined by the
bandwidth $W$ or the{\sc \ } Fermi energy $E_{F}$.

It was noted by Migdal himself \cite{M} that at $\vec{q}=0$ the{\sc \ }%
vertex corrections are not small (so they are important for optical
conductivity and Raman scattering). This case was considered in detail by
Engelsberg and Schrieffer \cite{ES}. Migdal's theorem is also violated for a
one-dimensional Fermi surface \cite{AM}.

Traditionally nonadiabatic corrections were considered to be small because
the Migdal parameter $\Omega _{ph}/\tilde{W}$ was small \cite{M,AGD,ES,S,AM}%
. But recently some materials (for example, fullerenes and high-$T_{c}$
superconductors) where $\Omega _{ph}\sim \tilde{W}$ were discovered. So
Migdal's theorem about the smallness of the nonadiabatic corrections \cite
{M,AGD,AM} can be violated. This gives grounds for not only discussing these
corrections, but even examining an antiadiabatic limit $\Omega _{ph}\gg 
\tilde{W}$ \cite{T}. But even with small $\Omega _{ph}/\tilde{W}$ there are
contradictory conclusions about the importance of such corrections in the
normal and superconducting state (the authors of Refs{\sc .}\cite{M,AGD,K}
consider them to be negligible, while the authors of \cite
{PSG1,GPS2,GPS3,T,DDLcjp,DDLpla,DDKOprb} think the opposite is true). The
most intriguing question is whether the violation of adiabaticity leads to
an increase of the interelectron interaction (and, as a result, to an
increase of $T_{c}$). There are contradictory results about the effect of
the vertex corrections on $T_{c}$. For example, Takada \cite{T} reported
some increase of $T_{c}$ for small parameters $\Omega _{ph}/E_{F}$. Even
more significant increase was claimed in \cite{GPS2}. At the same time, in
Refs.{\sc \ }\cite{CLX,KM,CDG}\cite{KM} a decrease of $T_{c}$ was obtained%
{\sc \ }due to the vertex corrections.

One can distinguish between the{\sc \ } two sorts of the nonadiabatic
corrections to the electron self-energy: 1) the vertex ones, resulted from
the higher order corrections of perturbation theory, as vertex function $%
\Gamma \neq 1$ even at an infinite electron (or hole) band $-\infty
<\varepsilon <+\infty $ \cite{ES,S,T}, and 2) the corrections due to the{\sc %
\ }finite band width,{\sc \ }$-W<\varepsilon <W$ \cite
{ES,BGPV,PSG1,GPS2,GPS3,T,K}.

Nonadiabatic corrections in the superconducting state enter the expression
for $T_{c}$ in two ways: indirectly through renormalization of $Z$ and
directly in the equation for the order parameter $\Delta $. At the same
time, analytical calculation of anomalous (crossing) diagrams is rather
difficult. Therefore before proceeding to calculations in the
superconducting state we want to clarify the role of such corrections in the
normal state. The main goal of this paper is to compare different approaches
to calculation of nonadiabatic corrections to the electron quasiparticle
self-energy in the normal state (or more precisely to the mass
renormalization factor $Z$). There are two methods to take nonadiabatic
corrections into account. One of them, which can be called Migdal's one, is
based on the solution to the Bethe-Salpeter equation for the vertex function 
\cite{M,ES,S} in the ladder approximation. In the lowest approximation, the
first correction to the unity vertex is determined by the diagram in Fig.1.
Then with the obtained vertex function the self-energy $\Sigma $ is
calculated \cite{DDLcjp,DDLpla,KM,PSG1}. This traditional method does not
allow to take into account higher-order diagrams analytically because of
complexity of the{\sc \ }integration over momenta.

In 1989 the{\sc \ }authors of the Ref.\cite{CLX} considered a{\sc \ }%
non-ladder approximation, using the Ward identity. Later Y.~Takada used the
same idea in his method \cite{T}, which he called the gauge-invariant
self-consistent (GISC) method. In this method, based on the Ward identity 
\cite{Sch}

\begin{eqnarray}
&&i\omega _{\nu }\Gamma (i\omega _{n},i\omega _{n}-i\omega _{\nu },\vec{k},%
\vec{k}-\vec{q})-\vec{q}\vec{\Gamma}(i\omega _{n},i\omega _{n}-i\omega _{\nu
},\vec{k},\vec{k}-\vec{q})  \nonumber \\
&=&G^{-1}(i\omega _{n},\vec{k})-G^{-1}(i\omega _{n}-i\omega _{\nu },\vec{k}-%
\vec{q}),  \label{eq1.5}
\end{eqnarray}
the vector term is neglected and the scalar vertex function is chosen to be
a functional of the self-energy, which is supposed to be independent from
momentum. Recently there appeared another method \cite{CDG}, aiming at
improving the GISC approach as it takes into account the{\sc \ }momentum
dependence of the vertex function. So, when comparing these methods with the%
{\sc \ }Migdal's one, we also discuss the{\sc \ }validity of these
assumptions. In fact, for the self-energy we calculate all the diagrams up
to, and including the second order. The parameter $\Omega _{ph}/W\equiv
m_{0} $ is taken to be small.

For simplicity, we consider the model of the Fermi liquid with the usual
electron-phonon Hamiltonian \cite{AM} and Einstein phonon \cite{boson}
spectrum with the{\sc \ }so-called Eliashberg spectral function $\alpha
^{2}F_{E}(\Omega )=\frac{1}{2}\lambda \Omega _{ph}\delta (\Omega -\Omega
_{ph})$, where $\Omega _{ph}$ is the phonon frequency independent from
momentum, and $\lambda $ is the electron-phonon interaction constant. It is
assumed that the density of states is constant $N(0)$ for $-W<\varepsilon <W$
where $2W$ is the bandwidth. The chemical potential is zero, which
corresponds to a half-filled band. It is assumed that $\lambda \sim 1$.

We are interested in the quasiparticle self-energy $\Sigma $ on the
imaginary axis which is determined by the diagram in Fig.1. The phonon
Green's function is $D_{0}(i\omega _{\nu })=-\Omega _{ph}^{2}/(\omega _{\nu
}^{2}+\Omega _{ph}^{2})$, where $\omega _{\nu }=2\nu \pi T$, and the
electron Green's function is $G(i\omega _{n},\vec{p})=(i\omega
_{n}-\varepsilon _{\vec{p}}-\Sigma (i\omega _{n},\vec{p}))^{-1}$, where $%
\omega _{n}=(2n+1)\pi T$, $\varepsilon _{\vec{p}}$ is the bare electron
spectrum.

The vertex function itself (two-point one) is only of academic interest. We
are, however, interested in the quasiparticle self-energy $\Sigma $ which
affects many physical observables. Assuming that $\Sigma (i\omega _{n},\vec{k%
})$ weakly depends on $\vec{k}$, we take $|\vec{k}|=p_{F}=const$, and then 
\begin{eqnarray}
\Sigma (i\omega _{n}) &=&\int\limits_{0}^{\infty }d\Omega \alpha
^{2}F_{E}(\Omega )T\sum\limits_{\omega _{\nu }}\frac{2\Omega }{\omega _{\nu
}^{2}+\Omega ^{2}}\frac{1}{N(0)}  \nonumber \\
&&\times \sum\limits_{\vec{q}}G(i\omega _{n}-i\omega _{\nu },\vec{k}-\vec{q}%
)\Gamma (i\omega _{n},i\omega _{n}-i\omega _{\nu },\vec{q}).  \label{eq1}
\end{eqnarray}

The methods differ in the choice of the vertex function $\Gamma $, and this
will be discussed later. One should note that $\Gamma (\vec{q})$ enters the
integral for $\Sigma ${\sc ,} and all $\vec{q}$'s contribute to it. For
simplicity we will obtain results for $T=0$ and the variables will be{\sc \ }%
changed as follows: $\omega _{n}\longrightarrow \omega $, $\omega _{\nu
}\longrightarrow \nu $, remaining on the imaginary axis.

\section{Migdal's method}

\label{SEmigdalpage}

In the method which can be called Migdal's, the first correction to the
unity vertex function $\Gamma ^{(1)}$ is considered. This is shown in Fig1.
It has been estimated in many papers \cite{M,AGD,S}, but now it is becoming
especially significant \cite{BGPV,PSG1,GPS2,GPS3}because the cases were
found where{\sc \ }as the parameter $\lambda \Omega _{ph}/W$ may be not
small.

The diagram in Fig.1 corresponds to the expression

\begin{eqnarray}
\Gamma ^{(2)}(i\omega _{n},i\omega _{n}-i\omega _{\nu },\vec{q})
&=&T\sum\limits_{\omega _{\nu ^{\prime }}}\sum\limits_{\vec{q}^{\prime
}}\int\limits_{0}^{\infty }d\Omega \frac{\alpha ^{2}F_{E}(\Omega )2\Omega }{%
N(0)(\omega _{\nu ^{\prime }}^{2}+\Omega ^{2})}  \nonumber \\
&&\times G^{(0)}(i\omega _{n}-i\omega _{\nu ^{\prime }},\vec{p}-\vec{q}%
^{\prime })G^{(0)}(i\omega _{n}-i\omega _{\nu ^{\prime }}-i\omega _{\nu },%
\vec{p}-\vec{q}^{\prime }-\vec{q}).  \label{eq3.5}
\end{eqnarray}

Assuming that $|\vec{p}-\vec{q}^{\prime }|\sim p_{F}$ and expanding $%
\varepsilon (\vec{p}-\vec{q}^{\prime }-\vec{q})\approx \varepsilon (\vec{p}-%
\vec{q}^{\prime })-qV_{F}\cos \theta $ we get for $T=0$

\begin{eqnarray}
\Gamma ^{(2)}(i\omega ,i\omega -i\nu ,\vec{q}) &=&\lambda \frac{1}{4}%
\int_{-1}^{1}d\eta \frac{1}{-im^{\prime }+Q\eta }  \nonumber \\
&&\times \{\ln \frac{1+im+\frac{1}{m_{0}}}{1-im+\frac{1}{m_{0}}}+\ln \frac{%
1-im}{1+im}+\ln \frac{1+im-im^{\prime }}{1-im+im^{\prime }}  \nonumber \\
&&+\ln \frac{1-im+im^{\prime }-Q\eta +\frac{1}{m_{0}}}{1+im-im^{\prime
}+Q\eta +\frac{1}{m_{0}}}\},  \label{1315}
\end{eqnarray}
where $m=\omega /\Omega _{ph}$, $m^{\prime }=\nu /\Omega _{ph}$, $%
Q=qV_{F}/\Omega _{ph}$, $m_{0}=\Omega _{ph}/W$.

This is a nonanalytical function at $\omega \rightarrow 0$, $\vec{q}%
\rightarrow 0$, i.e. the result depends on the order of taking the limits.
Let us define the{\sc \ }dynamical and static vertex functions

\begin{eqnarray}
\Gamma _{d} &=&\Gamma (\vec{q}=0,\nu \rightarrow 0,\omega );  \nonumber \\
\Gamma _{s} &=&\Gamma (\vec{q}\rightarrow 0,\nu =0,\omega ).  \label{1316}
\end{eqnarray}

From (\ref{1315}) one can get

\[
\Gamma _{d}^{M}=\Gamma _{M}(\vec{q}=0,\nu \rightarrow 0,\omega )=\lambda
\lbrack \frac{1}{1+(\frac{\omega }{\Omega _{ph}})^{2}}-\frac{1+\frac{W}{%
\Omega _{ph}}}{(1+\frac{W}{\Omega _{ph}})^{2}+(\frac{\omega }{\Omega _{ph}}%
)^{2}}], 
\]

\[
\Gamma _{s}^{M}=\Gamma _{M}(\vec{q}\rightarrow 0,\nu =0,\omega )=-\lambda 
\frac{1+\frac{W}{\Omega _{ph}}}{(1+\frac{W}{\Omega _{ph}})^{2}+(\frac{\omega 
}{\Omega _{ph}})^{2}}. 
\]

It is evident that the nonanalyticity mentioned above gives

\[
\Gamma _{d}^{M}-\Gamma _{s}^{M}=\lambda \frac{1}{1+(\frac{\omega }{\Omega
_{ph}})^{2}}. 
\]

At $V_{F}|\vec{q}|\ll |\nu |$

\begin{eqnarray}
&&\Gamma ^{(2)}(i\omega ,i\omega -i\nu ,\vec{q})  \nonumber \\
&=&\lambda \frac{\Omega _{ph}}{\nu }(\arctan \frac{\omega }{\Omega _{ph}}%
-\arctan \frac{\omega -\nu }{\Omega _{ph}}-\arctan \frac{\omega }{W+\Omega
_{ph}}  \nonumber \\
+\arctan \frac{\omega -\nu }{W+\Omega _{ph}}) &=&-\frac{\Sigma
^{(1)}(i\omega )-\Sigma ^{(1)}(i\omega -i\nu )}{i\nu },
\end{eqnarray}
which satisfies the Ward identity (\ref{eq1.5}).

At $V_{F}|\vec{q}|\gg |\nu |$ one has $\Gamma ^{(2)}(i\omega ,i\omega -i\nu ,%
\vec{q})\sim \lambda \Omega _{ph}/W,$ which complies with Refs. \cite{M,AGD}.

To calculate the self-energy we expand the last term of (\ref{1315}) in $Q$,
leaving the first two terms{\sc \ }as they are. Then

\begin{eqnarray}
\Gamma ^{(2)}(i\omega ,i\omega -i\nu ,\vec{q}) &=&\lambda \frac{\Omega _{ph}%
}{qV_{F}}\arctan \frac{qV_{F}}{\nu }  \nonumber \\
&&\times (\arctan \frac{\omega }{\Omega _{ph}}-\arctan \frac{\omega -\nu }{%
\Omega _{ph}}-\arctan \frac{\omega }{W+\Omega _{ph}}+\arctan \frac{\omega
-\nu }{W+\Omega _{ph}})  \nonumber \\
&&-\lambda \frac{\frac{\Omega _{ph}^{2}}{qV_{F}W}(1+\frac{\Omega _{ph}}{W})(%
\frac{qV_{F}}{\Omega _{ph}}-\frac{\nu }{\Omega _{ph}}\arctan \frac{qV_{F}}{%
\nu })}{[1+2\frac{\Omega _{ph}}{W}+(\frac{\Omega _{ph}}{W})^{2}(1+(\frac{%
\omega }{\Omega _{ph}})^{2}-2\frac{\omega \nu }{\Omega _{ph}^{2}}+(\frac{\nu 
}{\Omega _{ph}})^{2})]}.  \label{eq4}
\end{eqnarray}

\bigskip In Fig. 2a we show $\Gamma ^{(2)}(\nu )$ (numerical (\ref{1315})
and approximate analytical (\ref{eq4}) results) for the case $\omega =0,$ as
well as the{\sc \ }result of the numerical integration of Eq.(4). One can
see that there is a good agreement between these plots, so the expansion in
small $Q$ is well justified.

The self-energy (\ref{eq1}) consists then of three terms:

\begin{eqnarray}
\Sigma=\Sigma^{(1)}+\Sigma_v^{(2)}+\Sigma _r^{(2)}.
\end{eqnarray}

The first order term is \cite{ES} 
\begin{equation}
\Sigma ^{(1)}(i\omega )=-i\lambda \Omega _{ph}\arctan \frac{\omega }{\Omega
_{ph}}+i\lambda \Omega _{ph}\arctan \frac{\omega }{W+\Omega _{ph}}.
\label{s1}
\end{equation}

The vertex correction $\Sigma _v^{(2)}$ is obtained when one substitutes $%
G=G^{(0)}$ and $\Gamma =\Gamma^{(2)}$ (\ref{eq4}) into (\ref{eq1})

\begin{eqnarray}
\Sigma_v^{(2)}(i\omega )=\Sigma_v^{(2)a}(i\omega )+\Sigma_v^{(2)b}(i\omega ),
\label{d1.28}
\end{eqnarray}

\[
\Sigma _{v}^{(2)a}(i\omega )=i\lambda ^{2}\frac{\Omega _{ph}}{p_{F}V_{F}}%
d_{v}^{a}(m,m_{0},m_{0}^{\prime }), 
\]
$m=\omega /\Omega _{ph}$, $m_{0}=\Omega _{ph}/W$, $m_{0}^{\prime }=\Omega
_{ph}/(p_{F}V_{F})$.

\begin{eqnarray}
d_{v}^{a}(m,m_{0},m_{0}^{\prime }) =-\frac{1}{2\pi }\int\limits_{-\infty}^{+
\infty }\frac{dy}{y^{2}+1} \arctan \frac{1}{m_{0}(m-y)}  \nonumber \\
\times (\arctan m-\arctan (m-y)-\arctan \frac{m}{\frac{1}{m_{0}}+1}+ \arctan 
\frac{ m-y}{\frac{1}{m_{0}}+1})  \nonumber \\
\times (2\arctan \frac{2}{ym_{0}^{\prime }} -\frac{1}{2}ym_{0}^{\prime }\ln
( \frac{4}{y^{2}m{^{\prime }}_{0}^{2}}+1)).
\end{eqnarray}

\[
\Sigma _{v}^{(2)b}(i\omega )=i\lambda ^{2}\frac{\Omega _{ph}}{p_{F}V_{F}}%
d_{v}^{b}(m,m_{0},m_{0}^{\prime }), 
\]
where

\begin{eqnarray}
d_{v}^{b}(m,m_{0},m_{0}^{\prime })=-\frac{1}{2\pi }m_{0}^{2}(1+m_{0})
\int_{-\infty }^{+\infty }dm^{\prime }\frac{1}{m^{\prime 2}+1} \arctan \frac{
1}{m_{0}(m^{\prime }-m)}  \nonumber \\
\times \{\frac{2}{m{^{\prime }}_{0}^{2}}- \frac{2m^{\prime }}{m_{0}^{\prime
} }\arctan \frac{2}{m^{\prime}m_{0}^{\prime }} +\frac{m^{\prime }{}^{2}}{2}%
\ln [1+(\frac{2}{m^{\prime }m_{0}^{\prime }})^{2}]\}.
\end{eqnarray}

For $\omega /\Omega _{ph}\ll 1$, $\Omega _{ph}/W\ll 1$ it gives

\begin{eqnarray}
\Sigma _{v}^{(2)}(i\omega )=\omega \lambda ^{2}\frac{\Omega _{ph}}{p_{F}V_{F}%
}i[\frac{\pi ^{2}}{8}+(\frac{p_{F}V_{F}}{W})^{2}].  \label{591}
\end{eqnarray}

The plot of $\Sigma _{v}^{(2)}(i\omega )$ is shown in Fig.2b. The numerical
calculations (the{\sc \ } exact integration with Eq.(\ref{1315})) give
similar results for the{\sc \ }small $\omega $'s . The existing discrepancy
for larger frequencies is due to our approximation for $\Gamma $.

Another term is the so-called rainbow diagram. As the inner part of the
rainbow diagram corresponds to $\Sigma ^{(1)}(i\omega -i\nu )$, the second
order rainbow diagram equals

\begin{eqnarray}
\Sigma^{(2)}_r(i\omega)= \int\limits_0^\infty d\Omega \alpha ^2F_E(\Omega )
T\sum\limits_{\omega _\nu }\frac{2\Omega }{\omega _\nu ^2+\Omega ^2} \frac 1{%
N(0)}  \nonumber \\
\times \sum\limits_{\vec q}[G (i\omega _n-i\omega _\nu, \vec k-\vec q )]^2
\Sigma^{(1)}(i\omega_n-i\omega_{\nu}),  \label{eqd05}
\end{eqnarray}
which results in

\[
\Sigma _{r}^{(2)}(i\omega )=i\lambda ^{2}\Omega _{ph}d_{r}(m,m_{0}),%
\mbox{
where} 
\]

\begin{eqnarray}
d_{r}(m,m_{0}) &=&\frac{1}{m_{0}(m^{4}+2m^{2}+1+\frac{1}{m_{0}^{4}}+2(\frac{m%
}{m_{0}})^{2}-\frac{2}{m_{0}^{2}})}  \nonumber \\
&&\times (m\ln \frac{m^{2}+2^{2}}{m^{2}+(2+\frac{1}{m_{0}})^{2}}+(\arctan 
\frac{m}{2}-\arctan \frac{m}{2+\frac{1}{m_{0}}})(m^{2}-1+\frac{1}{m_{0}^{2}})
\nonumber \\
&&-2m\ln \frac{m_{0}+1}{m_{0}+2}).  \label{eqd09}
\end{eqnarray}

At $\omega /\Omega _{ph}\ll 1$, $\Omega _{ph}/W\ll 1$, it equals 
\begin{eqnarray}
\Sigma _{r}^{(2)}(i\omega )=i\frac{1}{2}\lambda ^{2}\omega \frac{\Omega
_{ph} }{W}.  \label{sigma_r}
\end{eqnarray}

For $\omega /\Omega _{ph}\ll 1$, $\Omega _{ph}/W\ll 1$, $\Omega
_{ph}/(p_{F}V_{F})\ll 1$ one can get

\[
Z_{M}\equiv 1-\frac{\Sigma (i\omega )}{i\omega }=1+\lambda -\frac{\pi ^{2}}{8%
}\lambda ^{2}\frac{\Omega _{ph}}{p_{F}V_{F}}-\frac{1}{2}\lambda ^{2}\frac{%
\Omega _{ph}}{W}-\lambda ^{2}\frac{\Omega p_{F}V_{F}}{W^{2}}. 
\]
One can see that the last expression has three terms with different
denominators. When, for example, $p_{F}V_{F}\sim \Omega _{ph}$, the main
contribution comes from the term $-\frac{\pi ^{2}}{8}\lambda ^{2}\Omega
_{ph}/p_{F}V_{F}$ (even for an infinite bandwidth $W$). In Fig.2b there are
shown for comparison analytical and numerical plots for the self energy
correction $\Sigma _{v}^{(2)}$. It is seen that the approximation used to
calculate the correction is well justified.

If one assumes $W=p_{F}V_{F}$, 
\begin{equation}
Z_{M}\approx 1+\lambda -2.7\lambda ^{2}\frac{\Omega _{ph}}{W}.  \label{592}
\end{equation}

One can see that these corrections lower the renormalization function. Had%
{\sc \ }this tendency persist in all higher orders, this would have{\sc \ }%
resulted in an{\sc \ }instability (see below).

\section{Cai, Lei, and Xie's approximation}

Following the{\sc \ }paper\cite{GS}, the authors{\sc \ }of Ref.\cite{CLX}
neglected the{\sc \ }momentum dependence of the vertex function, considering 
$q\gg p_{F}\nu /W$, and interpolated it between zero-frequency and
infinite-frequency limits:

\bigskip 
\[
\bigskip \Gamma _{CLX}^{(2)}(i\omega _{n},i\omega _{n^{\prime }})=(1+2\int
d\omega \frac{\omega \alpha ^{2}F(\omega )}{\omega ^{2}+(\omega _{n}-\omega
_{n^{\prime }})^{2}}\Lambda _{0}\frac{2(\Lambda _{\infty }/\Lambda _{0})}{%
\omega _{m}^{2}+\omega _{n}^{2}+2(\Lambda _{\infty }/\Lambda _{0})})^{-1}-1, 
\]

\bigskip where $\Lambda _{0}(\omega )=\frac{0.293\omega }{\omega +0.667W}$,
and $\Lambda _{\infty }(\omega )/\Lambda _{0}(\omega )=\frac{4\sqrt{2}}{3}%
\frac{0.667W+\omega }{0.586}.$

This results in an effective interaction

\[
V_{e-ph}(\omega -\omega ^{\prime })=\frac{1}{N(0)}\frac{\lambda \Omega ^{2}}{%
(\omega -\omega ^{\prime })^{2}+\Omega ^{2}}\times \frac{1}{1+\frac{\lambda
\Omega ^{2}}{(\omega -\omega ^{\prime })^{2}+\Omega ^{2}}\times \frac{%
0.293\Omega }{\Omega +0.667W}}. 
\]

Combined with the contribution from the rainbow diagram{\sc ,} which is the
same as in the Migdal method (\ref{sigma_r}), this gives

\[
Z_{CLX}\approx 1+\lambda -1.38\lambda ^{2}\frac{\Omega }{W}. 
\]

The non-ladder approximation which was also discussed in \cite{CLX} was
developed in detail by Takada, and this is considered below.

\section{Kostur and Mitrovi\'c's approximation}

There has been a series of papers, e.g. Ref. \cite{KM}, which considered the%
{\sc \ }effect of the{\sc \ }vertex corrections on $T_{c}$ not only for an
isotropic EPI, but also for spin (antiferromagnetic) fluctuations with a
pronounced scattering at the wave vector $\vec{Q}^{\ast }=(\pi ,\pi )$. In
the paper \cite{KM} the authors used the following equations for the
self-energy $\Sigma (\vec{k},i\omega _{n})=i\omega _{n}[1-Z_{n}(\vec{k})]$:

\begin{eqnarray}
i\omega _{n}Z_{n}^{(2)}(\vec{k}) &=&-\frac{T^{2}}{4W}\sum_{n^{\prime
}n^{\prime \prime }}\lambda (n-n^{\prime })\lambda (n-n^{\prime \prime }) 
\nonumber \\
&&\times \int_{-W}^{W}d\epsilon _{{\bf k}^{\prime }}\int_{-W}^{W}d\epsilon _{%
{\bf k}^{\prime \prime \prime }}\int_{-W}^{W}d\epsilon _{{\bf k}^{\prime
\prime }}M({\bf k,k}^{\prime \prime \prime }{\bf ;k}^{\prime }{\bf ,k}%
^{\prime \prime })\frac{i\omega _{n^{\prime }}i\omega _{n^{\prime \prime
}}i\omega _{n^{\prime \prime \prime }}}{Z_{n^{\prime }}Z_{n^{\prime \prime
}}Z_{n^{\prime \prime \prime }}}  \nonumber \\
&&\times \frac{1}{[i\omega _{n^{\prime }}^{2}-(\epsilon _{{\bf k}^{\prime
}}/Z_{n^{\prime }})^{2}][i\omega _{n^{\prime \prime }}^{2}-(\epsilon _{{\bf k%
}^{\prime \prime }}/Z_{n^{\prime \prime }})^{2}][i\omega _{n^{\prime \prime
\prime }}^{2}-(\epsilon _{{\bf k}^{\prime \prime \prime }}/Z_{n^{\prime
\prime \prime }})^{2}]},  \label{416}
\end{eqnarray}
where $\lambda (m)=\int_{0}^{\infty }d\Omega \alpha ^{2}F(\Omega )\frac{%
2\Omega }{\nu _{m}^{2}+\Omega ^{2}}$, and $M({\bf k,k}^{\prime \prime \prime
}{\bf ;k}^{\prime }{\bf ,k}^{\prime \prime })$ is a geometrical factor being
a complicated function of momenta which is considered in detail in Ref.\cite
{KM}. For an isotropic EPI and a three-dimensional spherical Fermi-surface
it is approximated by its value on the Fermi surface $M({\bf k,k}^{\prime
\prime \prime }{\bf ;k}^{\prime }{\bf ,k}^{\prime \prime })=1$. It gives

\[
Z_{n}=1+\frac{\pi T}{|\omega _{n}|}\sum_{n^{\prime }}\lambda (n-n^{\prime
})\Gamma _{KM}(n,n^{\prime })s_{n}s_{n^{\prime }}a_{n^{\prime }},%
\mbox{
where} 
\]
$\lambda (m)=\lambda \frac{\Omega _{ph}^{2}}{\nu _{m}^{2}+\Omega _{ph}^{2}}$%
, $a_{n}=2/\pi \arctan (W/Z_{n}|\omega _{n}|)$, $s_{n}=\mbox{sign }\omega
_{n}$, and correction to the vertex function equals

\begin{eqnarray}
\Gamma _{KM}^{(2)}(n,n^{\prime })=-\frac{\pi ^{2}T}{2W}\sum_{n^{\prime
\prime }}\lambda (n-n^{\prime \prime })s_{n^{\prime }+n^{\prime \prime
}-n}s_{n^{\prime \prime }}a_{n^{\prime }+n^{\prime \prime }-n}a_{n^{\prime
\prime }}.  \label{418}
\end{eqnarray}

One can find that

\[
\Gamma _{KM}^{(2)}(n,n^{\prime })=1-\lambda \frac{\pi }{4}\frac{\Omega _{ph}%
}{W}(\pi -2|\arctan \frac{\omega _{n^{\prime }}}{\Omega _{ph}}-\arctan \frac{%
\omega _{n}}{\Omega _{ph}}|) 
\]
and so does not depend from $q$. However, for $q\sim p_{F}$ which give the
main contribution to the self-energy it presents the correct order of the
correction to the vertex function (compare with Eq.(\ref{eq4})) and the
self-energy. From Eq. (2) one has for $\omega /\Omega _{ph}\ll 1$, $\Omega
_{ph}/W\ll 1$

\[
Z_{KM}=1+\lambda +\lambda ^{2}\frac{\Omega _{ph}}{W}(-\frac{\pi ^{2}}{8}-%
\frac{1}{2})\approx 1+\lambda -1.73\lambda ^{2}\frac{\Omega _{ph}}{W}. 
\]

This is similar to Eq. (\ref{592}), but has a different numerical
coefficient, due to neglecting the $\vec{q}$-dependence in the vertex
function (\ref{418}).

\section{C. Grimaldi, L. Pietronero, and S. Str\"{a}ssler's approximation}

In the paper \cite{PSG1} the authors started with the same equation as in
Migdal method, but made a series of approximate assumptions (for example,
expanded in small $q$'s to take integrals) and gave the following estimate
for the vertex function:

\begin{eqnarray}
\Gamma _{GPS}^{(2)}(\omega ,Q,\Omega ,W) &=&\frac{\lambda }{Q}[[\arctan
m-\arctan \frac{m}{\frac{1}{m_{0}}+1}]\arctan \frac{Q}{m}  \nonumber \\
&&-[Q-m\arctan \frac{Q}{m}]\frac{(\frac{1}{m_{0}}+1)[(\frac{1}{m_{0}}%
+1)^{2}+2m^{2}]}{[(\frac{1}{m_{0}}+1)^{2}+m^{2}]^{2}}].
\end{eqnarray}
Here they set one of the external electronic frequencies to zero. In this
case this result is similar to that in the{\sc \ }Migdal's approximation (%
\ref{eq4}). However, to calculate $\Sigma $ one needs dependences on both
external frequencies.

There are some indications (see, e.g. \cite{Kulic}) that for a{\sc \ }small
hole doping strong Coulomb correlations renormalize the EPI, giving rise to
the strong forward (small-$q$) scattering peak, while the backward
scattering is strongly suppressed. With this idea in mind, in Ref. \cite
{GPS2} the electron-phonon coupling constant was assumed to have a cut-off
in a momentum space $|g_{\vec{p},\vec{k}}|^{2}=g^{2}(2k_{F}/q_{c})^{2}\theta
(q_{c}-|\vec{p}-\vec{k}|)$. An approximation for $q\rightarrow 0$ was used 
\cite{cut-off} and then $q_{c}=Q_{c}\cdot 2p_{F}$ was set to $q_{c}=2p_{F}$
giving the following result \cite{GCP} for $Z$:

\[
Z_{CGPS}\approx 1+\lambda -\frac{\pi }{4}\lambda ^{2}\frac{\Omega _{ph}}{W}%
\frac{1}{Q_{c}^{2}}. 
\]

For $Q_c=1$ it becomes

\[
Z_{CGPS}\approx 1+\lambda -0.8\lambda ^{2}\frac{\Omega _{ph}}{W}, 
\]
which gives the correct order of the correction to the self-energy , but
underestimates it (compare with Eq.(\ref{592})). Probably this is due to the%
{\sc \ }expansion in small $q$'s.

\label{W}

\section{Takada's GISC method}

\label{Wtakadapage}

According to Takada's gauge-invariant self-consistent (GISC) method, the
vertex function can be chosen as a functional of the self-energy \cite{T} $%
\Gamma _{T}=\Gamma _{T}[\Sigma _{T}]$. This choice is based on the Ward
identity \cite{ES} (\ref{eq1.5}) valid for all $\omega _{\nu }$'s and $\vec{q%
}$'s. In Ref. \cite{T} Takada proposes neglecting the{\sc \ }vector term in
the Ward identity and choosing \cite{N} 
\[
\Gamma _{T}[\Sigma ]=\Gamma _{T}(i\omega _{n},i\omega _{n}-i\omega _{\nu
})=1+(-\frac{\Sigma _{T}(i\omega _{n})}{2i\omega _{n}}-\frac{\Sigma
_{T}(i\omega _{n}-i\omega _{\nu })}{2(i\omega _{n}-i\omega _{\nu })}), 
\]
which corresponds to the estimates of Refs. \cite{ES,S} in this limit. If
this method worked it would allow one to calculate also superconducting
diagrams without complicated integrations over momenta.

The set of equations (1,2) is suggested to be solved by iterations. At the
first step $G=G^{(0)}$, $\Gamma =\Gamma _{T}^{(1)}=1$ give $\Sigma
_{T}=\Sigma _{T}^{(1)}=\Sigma ^{(1)}$ from Eq.(2). At the second step $%
G=G[\Sigma _{T}^{(1)}]$ ; and the correction $\Gamma _{T}^{(2)}$ for $\Omega
_{ph}/W\ll 1$ gives \cite{DDLpla}

\begin{eqnarray}
\Sigma _T^{(2)}(i\omega ) =-i\lambda ^2\Omega _{ph}\left\{ 
\begin{array}{r}
(\frac 14+\frac{\pi ^2}{16}) \frac \omega {\Omega _{ph}}, \mbox{ } | \omega
| \ll\Omega _{ph}, \\ 
\frac{\pi ^2}4\frac{\Omega _{ph}}\omega , \mbox{ } | \omega | \gg \Omega
_{ph}
\end{array}
\right. \ .  \label{eq3}
\end{eqnarray}

For small frequencies it results in 
\[
Z_{T}=1+\lambda +\lambda ^{2}(\frac{\pi ^{2}}{16}+\frac{1}{4})\approx
1+\lambda +0.9\lambda ^{2}, 
\]
which has the{\sc \ }incorrect sign and order of magnitude of the vertex
correction (compare with (\ref{592})).

During this calculation, the correction to the vertex $\Gamma _{T}^{(2)}$ is
of the order of $\lambda $ for any $q$, which contradicts the{\sc \ }
Migdal's theorem \cite{M} and can be valid only for $q\sim 0$ (see e.g. Ref. 
\cite{ES}). So in Takada's method the correction $\Sigma _{T}^{(2)}$ does
not have the{\sc \ }small parameter $\Omega _{ph}/W$. This can be explained
by the fact that the region $V_{F}|\vec{q}|\ll |\nu |$ where $\Gamma \sim
\lambda $ gives in fact a small contribution to $\Sigma $. If, on the other
hand, $V_{F}|\vec{q}|\gg |\nu |$, the vertex $\Gamma $ is of the order of $%
\lambda \Omega _{ph}/W$, which gives $\Sigma ^{(2)}\sim \lambda \Omega
_{ph}^{2}/W$ (see (\ref{591})).

To explain this result, one can find the vector vertex function $\vec{\Gamma}
$ in the lowest approximation and show that it really cannot be neglected.
The diagram in \mbox{Fig. 1} corresponds to $\vec{\Gamma}^{(2)}$ if $\Gamma
^{(1)}$ is substituted by $\vec{\Gamma}^{(1)}$ and $\Gamma ^{(2)}(i\omega
_{n},i\omega _{n}-i\omega _{\nu },\vec{q})$ - by $\vec{\Gamma}^{(2)}(i\omega
,i\omega -i\nu ,-\vec{q})$. Using electron-hole symmetry, in our model $\vec{%
\Gamma}(i\omega ,i\omega -i\nu ,\vec{k},\vec{k}-\vec{q})=\vec{\Gamma}^{(1)}+%
\vec{\Gamma}^{(2)}\mbox{, where }\vec{\Gamma}^{(1)}\approx \frac{\vec{k}}{m}%
, $

\begin{eqnarray}
\vec{q}\vec{\Gamma}^{(2)}(i\omega ,i\omega -i\nu ,\vec{q}) &=&\lambda \frac{1%
}{4}\int_{-1}^{1}d\eta \frac{\Omega Q\eta }{-im^{\prime }+Q\eta }  \nonumber
\\
&&\times \{\ln \frac{1+im+\frac{1}{m_{0}}}{1-im+\frac{1}{m_{0}}}+\ln \frac{%
1-im}{1+im}+\ln \frac{1+im-im^{\prime }}{1-im+im^{\prime }}  \nonumber \\
&&+\ln \frac{1-im+im^{\prime }-Q\eta +\frac{1}{m_{0}}}{1+im-im^{\prime
}+Q\eta +\frac{1}{m_{0}}}\}.  \label{1323}
\end{eqnarray}

Neglecting the $Q$-dependence in the last term of (\ref{1323}) one can get

\begin{eqnarray}
\vec{q}\vec{\Gamma}^{(2)}=-i\lambda \Omega _{ph}(1-\frac{\nu }{qV_{F}}%
\arctan \frac{qV_{F}}{\nu })(\arctan \frac{\omega }{\Omega _{ph}}-\arctan 
\frac{\omega -\nu }{\Omega _{ph}}).
\end{eqnarray}

We can find regions where scalar or vector terms in the Ward identity
dominate.

If one uses at $\Omega _{ph}/W\ll 1$ the expression (\ref{eq4}) for $\Gamma
^{(2)}$, one can find that the terms $\nu \Gamma ^{(2)}$ and $\vec{q}\vec{%
\Gamma}^{(2)}$ from (\ref{eq1.5}) are of the same order if $\nu /qV_{F}=0.43$%
. If, however, $\nu /qV_{F}<0.43$, the vector term dominates and cannot be
neglected.

\section{F. Cosenza, L. De Cesare, and M. Fusco Girard's improvement of the
GISC method}

Taking into account a{\sc \ }criticism of the GISC method \cite{DDLpla}, F.
Cosenza, L. De Cesare, and M. Fusco Girard \cite{CDG} tried to improve the%
{\sc \ }GISC method and did not neglect the vector term in the Ward
identity. They employed the simplest choice of the solution to the Ward
identity \cite{Ko}

\begin{equation}
\Gamma _{CDG}^{(2)}(\vec{k},i\omega _{n},\vec{k^{\prime }},i\omega
_{n^{\prime }})\approx \frac{(i\omega _{n^{\prime }}-i\omega _{n})[\Sigma (%
\vec{k},i\omega _{n})-\Sigma (\vec{k^{\prime }},i\omega _{n^{\prime }})]}{%
(i\omega _{n^{\prime }}-i\omega _{n})^{2}-|\alpha (\vec{k},\vec{k^{\prime }}%
)|^{2}|\vec{k}-\vec{k^{\prime }}|^{2}},  \label{cdg}
\end{equation}
where $\alpha (\vec{k},\vec{k^{\prime }})=(\vec{q}/2+\vec{k})/m\approx \vec{%
V_{F}}$. With $\Sigma (\vec{k},i\omega _{n})$ taken from Eq.(\ref{s1}) one
gets for $|\vec{q}|=|\vec{k}-\vec{k^{\prime }}|\sim p_{F}$

\[
\Gamma _{CDG}^{(2)}(\vec{k},i\omega _{n},\vec{k^{\prime }},i\omega
_{n^{\prime }})\sim \lambda (\frac{\Omega _{ph}}{W})^{2} 
\]
which is different from the Migdal's estimation \cite{M} $\Gamma ^{(2)}\sim
\lambda \frac{\Omega _{ph}}{W}$ (see (\ref{eq4})).

Substituting approximation (\ref{cdg}) into the equation for the self-energy
(\ref{eq1}) one can get 
\[
\Sigma _{CDG}^{(2)}(i\omega \rightarrow 0)\approx 0.25i\lambda ^{2}\omega (%
\frac{\Omega _{ph}}{W})^{2}(\ln \frac{W}{\Omega _{ph}})^{2} 
\]
which is different from (\ref{591}) and gives 
\[
Z_{CDG}\approx 1+\lambda -0.25\lambda ^{2}(\frac{\Omega _{ph}}{W})^{2}(\ln 
\frac{W}{\Omega _{ph}})^{2}. 
\]

This can be explained by the fact that the Ward identity is a{\sc \ }one
equation for two functions (the scalar and vector vertex functions) and thus
allows for multiple solutions \cite{Ko}. So there is no particular
preference in using (\ref{cdg}) instead of any other solution.

\section{Discussion}

\label{Wresultspage} \label{SEresultspage}

So in the paper we considered consistently the contribution of nonadiabatic
effects to the electron self-energy in the normal state at $T=0$ in second
orders of the EPI constant $\lambda $. Several methods of taking into
account nonadiabatic corrections were compared which can be summarized in
the Table ($p_{F}V_{F}=W$).

\bigskip The results following from the approximations of V. N. Kostur and
B. Mitrovi\'{c} \cite{KM}, E. Cappelluti, C. Grimaldi, L. Pietronero, and S.
Str\"{a}ssler \cite{PSG1,GPS2,GPS3,GCP} give correct estimations of the
order of magnitude and sign of the nonadiabatic corrections to the
self-energy. The analytical results were compared with numerical
calculations.

It is shown that the GISC method \cite{T} and its generalization \cite{CDG}
give for the self-energy overestimated and underestimated results
respectively. This is connected with the fact that the leading contribution
to $\Sigma $ comes from{\sc \ }the region $qv_{F}\gg \nu $ where $\Gamma
\sim \lambda \Omega _{ph}/W$. This means that in the Ward identity one may
not neglect the vector term as proposed by Takada. In the framework of the
standard (Migdal's) approach, in the lowest order, there is a calculable{\sc %
\ } parameter $\nu /qv_{F}$ ( the ratio of the frequency of an incoming
phonon to its momentum) for which the contributions of the vector and scalar
terms become of the same order. They are equal when $\nu /qv_{F}=0.43$. So
for the{\sc \ }most metals, where the main contribution to the self energy
comes from the region $\nu /qv_{F}\ll 1$, the GISC method may not be
applied. One should also be careful when choosing a{\sc \ }solution of the
Ward identity, so{\sc \ }as not to underestimate the scalar term, like in
Ref. \cite{CDG}. It seems that GISC method can work in systems with long
range interaction, for example, in doped semiconductors.

The difference in the renormalization factors $Z$ in the Table are due with
the different approximations used in the calculations of the momentum and
frequency dependence of vertex function. We can illustrate this in Fig. 3
where we represent the results for $\Gamma ^{(2)}$ in the Takada's ($q$%
-independent) approximation\cite{T}, the Migdal approach for different $q$
(Eqs. (4,27) of the present paper), and the $q$-independent Kostur,
Mitrovi\'{c} vertex function. We see that if the former overestimates the
vertex corrections, the latter underestimates them.

For a stable system the condition $%
\mathop{\rm Im}%
\Sigma (i\omega _{n})<0$ for $\omega _{n}>0$ must be satisfied \cite{DL} (it
is equivalent to $Z(i\omega _{n})>1$). However, as one can see from the
Table, the nonadiabatic corrections reduce{\sc \ }the renormalization
function $Z$, favoring the tendency towards instability. The critical $%
\lambda _{crit}$ at which it takes place (see Eqs. (10-16)) is given by

\[
\lambda _{crit}=\frac{\arctan m-\arctan \frac{m}{1/m_{0}+1}}{%
m_{0}d_{v}(m,m_{0})+d_{r}(m,m_{0}))}, 
\]
for $m\rightarrow 0$, where $d_{v}(m,m_{0})$ and $d_{r}(m,m_{0})$ were
defined above. The dependence of $\lambda $ on the parameter $\Omega _{ph}/W$
is shown in Fig.4. Thus, for a stable system there are constraints on $%
\lambda $ for a fixed Migdal parameter\cite{critical}. This can explain, for
example, the existence of PbBi with $\lambda \sim 3$. At the same time, in
some papers, e.g. \cite{AK}, it is stated that $\lambda \sim 1$
independently from the{\sc \ }Migdal parameter. As one sees from 
\mbox{Fig.
4}, it may be the case only for $\Omega _{ph}\sim W$.

This possible violation of the{\sc \ }analytical properties of the{\sc \ }%
one-particle Green's function could result in corresponding nonanaliticity
of the{\sc \ } two-particle Green's function, and thus in a charge response
function. This can lead to a charge instability.

It follows from our analysis that different approximations to the vertex
function give in the normal state quantitatively (and sometimes even
qualitatively!) different estimates{\sc \ }for the{\sc \ }self-energy. This
makes the{\sc \ }conclusions of these theories doubtful, as the same
approaches were used there to calculate $T_{c}$ in the superconducting
state. This means that only direct Migdal-type calculations can give an
answer to the question: Do nonadiabatic effects enhance critical temperature
of the superconducting transition? \cite{DD}

\section{Acknowledgment}

We thank T. Dahm, O. Gunnarsson, M.L. Kuli\'{c}, V.V. Losyakov, I.I. Mazin,
E.G. Maksimov, and N. Schopohl for helpful discussions.

\newpage

\bigskip Table

\begin{tabular}{|l|l|}
\hline
Method: & $Z^{(2)}$ \\ \hline\hline
Migdal (numerical), with $\Gamma ^{(2)}$ from Eq.(4) & $-2.7\lambda
^{2}\Omega _{ph}/W$ \\ \hline
Migdal (analytical), with $\Gamma ^{(2)}$ from Eq.(7) & $-2.7\lambda
^{2}\Omega _{ph}/W$ \\ \hline
CLX, Ref.\cite{CLX} & $-1.38\lambda ^{2}\Omega _{ph}/W$ \\ \hline
KM, Ref.\cite{KM} & $-1.73\lambda ^{2}\Omega _{ph}/W$ \\ \hline
CGPS, Ref.\cite{GPS2} & $-0.8\lambda ^{2}\Omega _{ph}/W$ \\ \hline
Takada's GISC, Ref.\cite{T} & +$0.9\lambda ^{2}$ \\ \hline
CDG, Ref.\cite{CDG} & $-0.25\lambda ^{2}(\Omega _{ph}/W)^{2}(\ln W/\Omega
_{ph})^{2}$ \\ \hline
\end{tabular}

\newpage

\centerline {\Large Figure Captions}

Fig.1. Equation for the vertex function $\Gamma ^{(2)}$ in Migdal's method.
The outgoing lines are shown for clarity and are not included in the
definition of $\Gamma $.

Fig. 2. a) Plot of $\Gamma ^{(2)}(\omega =0,i\nu )$ (Eq.(4), dashed line)
and $\Gamma _{appr}^{(2)}(\omega =0,i\nu )$ (Eq.(7), solid line) for $%
\lambda =1,\Omega _{ph}/W=0.1,\omega =0$, $q/p_{F}=0;0,2;0,4;0,6;0,8$. b)
Corresponding frequency dependence of $\Sigma _{v}^{(2)}(i\omega )$ (dashed
line) and $\Sigma _{v,appr}{}^{(2)}(i\omega )$ (solid line).

Fig.3. $\Gamma ^{(2)}(\omega =0,\nu )$in the Takada's approximation\cite{T}
(long dashed line); the Migdal approach for $q/p_{F}=0,0.5,1$ (solid lines,
from top to bottom) ; and the Kostur, Mitrovi\'{c} \cite{KM} vertex function
( short dashed line ).

Fig. 4. {$\lambda _{crit}$ at which $\Sigma (i\omega )=0$ vs the parameter $%
\Omega _{ph}/W.$}


\begin{references}
\bibitem{boson}  Formally, our conclusions are applicable to other low
frequency boson-like excitations, e.g. spin-fluctuations.
\bibitem{cut-off}  In fact the authors used a cut-off in angles and not in
momenta. Scattering by small momenta results instead in a nonexponential
behavior of $T_{c}$ as a function of the coupling constant \cite
{DDKOijmpb,DDKOprb}.
\bibitem{critical}  We should note that these values of $\lambda _{crit}$
are just the rough estimates, because our calculations were performed in the
first order in $\lambda .$
\end{references}
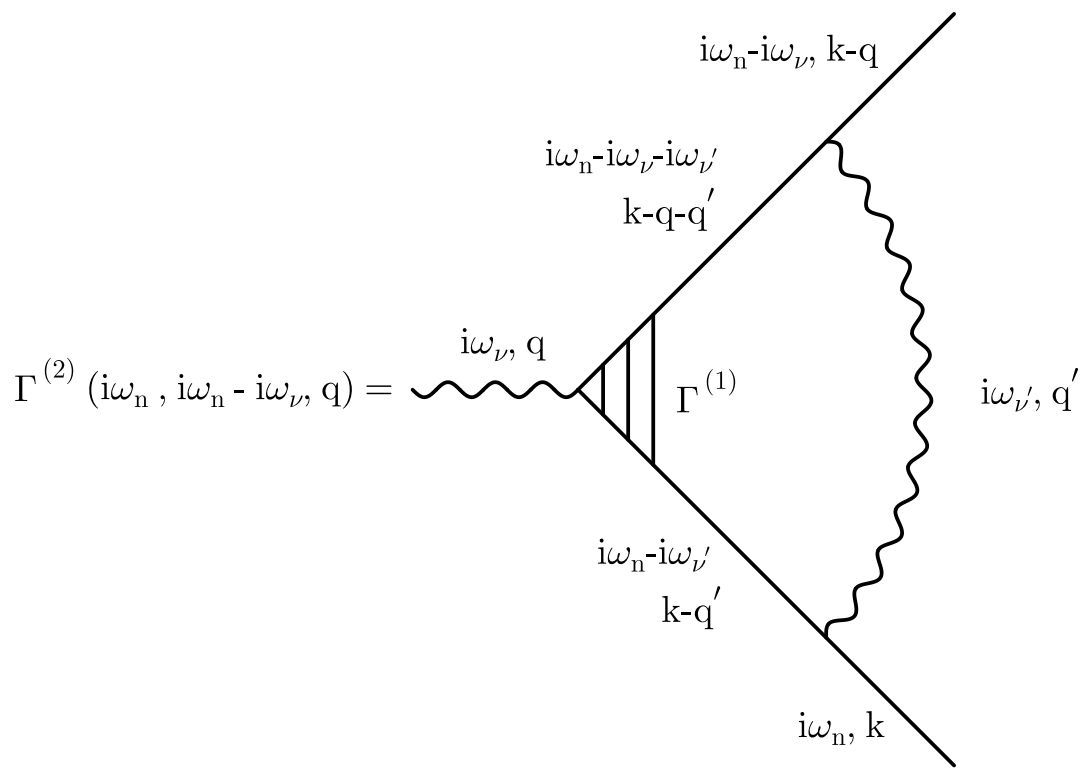
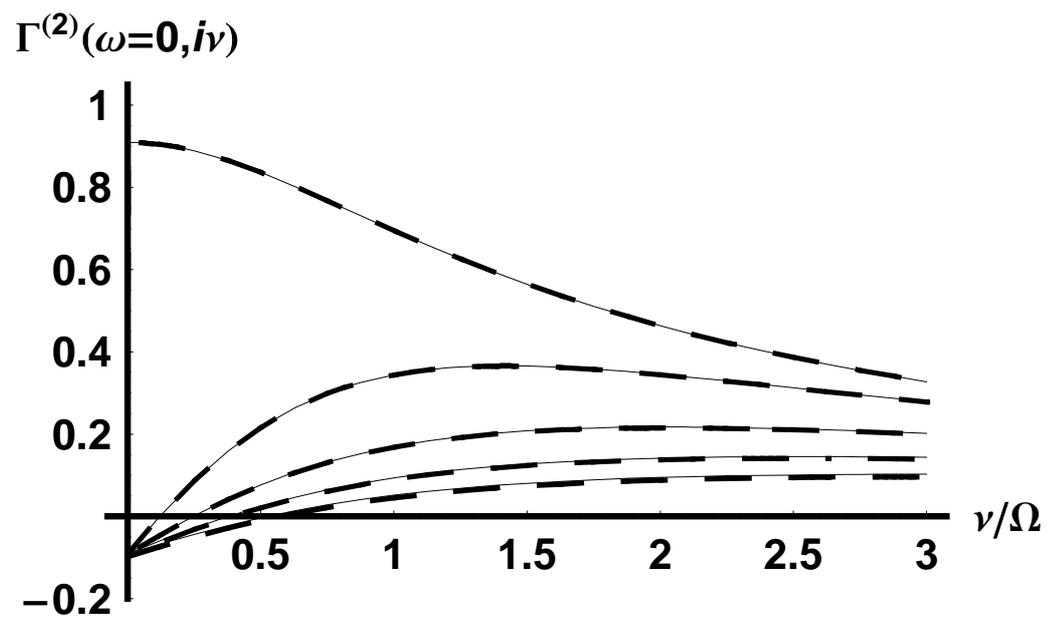
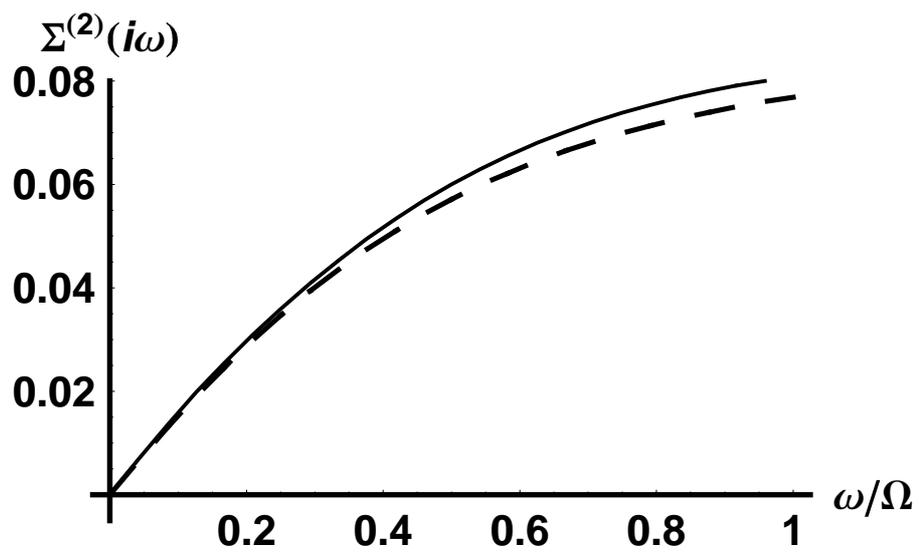
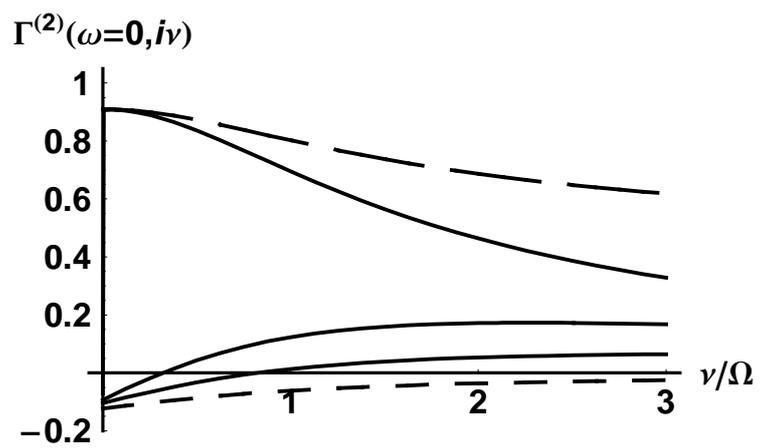
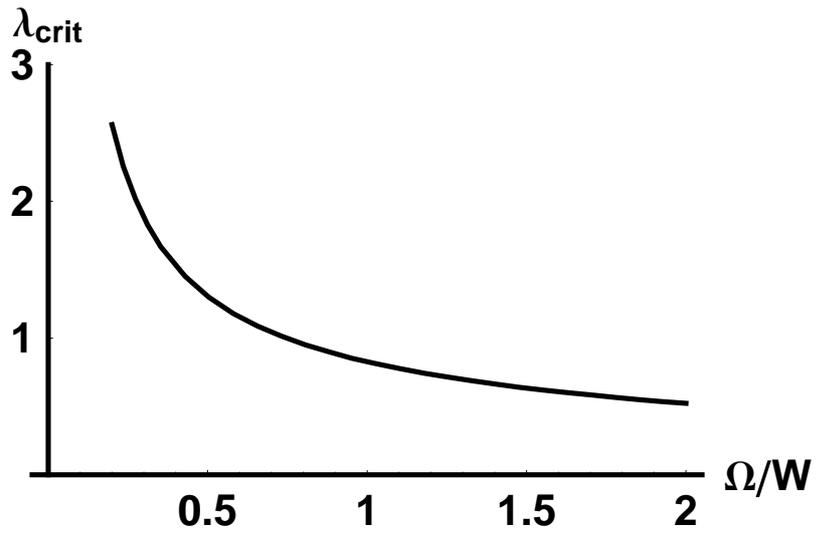
\end{document}